\documentclass[conference]{IEEEtran}

\usepackage{cite}
\usepackage{graphicx}
\usepackage{url}
\usepackage{multirow}
\usepackage[labelfont=bf]{caption}
\usepackage{supertabular}
\usepackage{relsize}
\usepackage{booktabs}
\usepackage{rotating}
\usepackage{paralist}
\usepackage{flushend}
\usepackage{ifthen}
\usepackage[table]{xcolor}

\newboolean{TechReport}
\setboolean{TechReport}{true}

\usepackage{color}

\newcommand{\new}[1]{\textcolor{black}{#1}} 

\begin{document}
%
\title{A Quality Framework for Agile Requirements:\\ A Practitioner's Perspective}


 
\author{\IEEEauthorblockN{Petra Heck}
\IEEEauthorblockA{Fontys Applied University\\
Eindhoven, The Netherlands\\
Email: p.heck@fontys.nl}
\and
\IEEEauthorblockN{Andy Zaidman}
\IEEEauthorblockA{Delft University of Technology\\
Delft, The Netherlands\\
Email: a.e.zaidman@tudelft.nl}
}


\maketitle

\begin{abstract}
Verification activities are necessary to ensure that the requirements are specified in a correct way. However, until now requirements verification research has focused on traditional up-front requirements. Agile or just-in-time requirements are by definition incomplete, not specific and might be ambiguous when initially specified, indicating a different notion of `correctness'. We analyze how verification of agile requirements quality should be performed, based on literature of traditional and agile requirements. This leads to an agile quality framework, instantiated for the specific requirement types of feature requests in open source projects and user stories in agile projects. We have \new{performed an initial qualitative validation of }our framework for feature requests with eight practitioners from the Dutch agile community, receiving overall positive feedback.   
\end{abstract}




\section{Introduction}

It is increasingly uncommon for software systems to be fully specified before implementation begins~\cite{ErnstCAISE2012}. As stated by Ernst et al.~\cite{Ernst2014}, ``The `big design up front' approach is no longer defensible, particularly in a business environment that emphasizes speed and resilience to change''. They observe that increasingly more industry projects treat requirements as tasks, managed with task management tools like Jira or Bugzilla. A similar task-based approach is seen in the agile movement (with e.g. user stories) and in open source projects\new{\cite{Warsta, Koch}}. In an earlier paper Ernst and Murphy~\cite{Ernst2012} use the term `just-in-time requirements' for this type of requirements. They observed that requirements are ``initially sketched out with simple natural language statements'', only to be fully elaborated (not necessarily specified in written form) when being developed. \new{In their analysis on requirements engineering for agile development Paetsch et al.~\cite{Paetsch} conclude that ``agile methods tend to err on the side of producing not enough documentation while traditional approaches tend to overdocument. \ldots As all agile approaches include at least a minimum of documentation, it is the responsibility of the development team to ensure enough documentation is available for future maintenance.''. By documentation they mean requirements documentation. According to a recent publication by IREB (International Requirements Engineering Board)~\cite{Grau} the documentation formats used in waterfall and agile environments do not differ that much and there is a continuous evolution in the definition of quality attributes for requirements. However, they do not explicitly define the current quality attributes for agile environments. This leads us to investigating the quality attributes for just-in-time requirements documentation in this paper.}  

\new{Both the\emph{ Business Analysis Body of Knowledge (BABOK)} guide~\cite{BABOK} and it's agile extension~\cite{BABOKagile} state that requirements should be verified before they are validated with stakeholders. In our paper we describe a \emph{verification} framework for agile requirements, thereby following the definitions of BABOK: ``Requirements verification ensures that requirements specifications and models meet the necessary standard of quality to allow them to be used effectively to guide further work. Requirements validation ensures that all requirements support the delivery of value to the business, fulfill its goals and objectives, and meet a stakeholder need.''}

Verification activities ensure that the requirements are specified in a correct way. Standards such as IEEE-830~\cite{IEEE830} define what `correct' means: requirements should be complete, unambiguous, specific, time-bounded, consistent, etc. However, this standard focuses on traditional up-front requirements. These are requirements sets that are completely specified before the start of design and development. The requirements set is signed-off by the stakeholders and considered as a contract of what has to be developed.

As agile or just-in-time requirements are `sketches' of what needs to be done they are by definition incomplete, not specific and might be ambiguous when initially specified. This indicates that the notion of quality for agile requirements is different from the notion of quality for traditional up-front requirements. The question is: \emph{how different}? 

We have not found a practical implementation of verification for agile requirements quality. This leads us to our main \textbf{research question}: \emph{How should we verify the quality of agile or just-in-time requirements?} This question assumes that correctly specified agile requirements contribute to a higher final software product quality, an assumption that has been considered to hold for traditional requirements, e.g.~\cite{Genova, Knauss, Karnata}. 

For agile and open source projects there is a body of related work both on Requirements Engineering (e.g.~\cite{Paetsch, Grau, Cleland2009, Noll2010, Cao, Waldmann}) and Quality Assurance (e.g.~\cite{Michlmayr, Aberdour, Huo, Klassen}). We did not find any specific related work on quality criteria for agile or open source requirements. 

In order to come to a framework for quality criteria for agile or open source requirements, we decided to start working from an existing quality framework for traditional requirements~\cite{HeckLaQuSo}. This framework provides an organization structure for verification criteria. While the framework was originally designed for software systems with traditional up-front requirements, we see it as a good starting point for our analysis of agile requirements as it is based on an extensive literature review of requirements verification. For this paper we focus on feature requests and user stories analog to the findings of Ernst~\cite{Ernst2012} that these are the common types of just-in-time requirements. 
This leads us to the detailed research questions:

\begin{itemize}
\item[\textbf{{[}RQ1a]}] Which quality criteria from literature on traditional up-front requirements do also apply to the verification of feature requests and user stories? 
\item[\textbf{{[}RQ1b]}] Do we see additional quality criteria for the verification of feature requests and user stories?
\end{itemize}

As the first version of our quality framework is based on literature and our own experience we deem it necessary to qualitatively validate the framework with practitioners. This leads to the following research question:
\begin{itemize}
\item[\textbf{{[}RQ2]}] How do practitioners value our list of quality criteria with respect to usability, completeness and relevance for the verification of feature requests?
\end{itemize}

The remainder of this paper is structured as follows. Section~\ref{Framework} explains the quality framework used. Section~\ref{FR} instantiates the quality framework as verification framework for feature requests. Section~\ref{US} explains what would be the difference if the framework is instantiated for user stories. Section~\ref{IQ} lists quality criteria that are automatically covered by the use of electronic tools. Sections~\ref{sec_es} and~\ref{sec_res} describe the experiment we performed with eight practitioners. Section~\ref{sec_dis} discusses the research questions, while Section~\ref{RelWork} highlights related work. Section~\ref{Conclusion} concludes this paper.

\section{A Quality Framework}\label{Framework}
\subsection{A Software Product Certification Model} \label{sec_SPCM}
As a basis for our agile verification framework we use the Software Product Certification Model (SPCM)~\cite{HeckLaQuSo}. This framework was introduced in 2010 and  is based on extensive literature research for traditional up-front requirements engineering. We will analyze which concepts are also applicable for our agile quality framework.

The SPCM divides a software product into six Product Areas, namely the \emph{Context Description}, which describes the environment of the system, the \emph{User Requirements}, the \emph{High-Level Design}, the \emph{Detailed Design}, the \emph{Implementation} and the \emph{Tests}. Each area is further divided into subparts, which are called elements. These elements can be separate artifacts, a chapter within a document, or different parts of a model.

Next to this division, the SPCM also defines specific certification (= verification) criteria for each area. These Specific Criteria (SC) are derived from three high-level Certification Criteria (CC):


\begin{description}
\item[\textit{{[}CC1{]}}] Completeness. All required elements in the Product Area should be present.  
\item[\textit{{[}CC2{]}}] Uniformity. The style of the elements in the Product Area should be standardized.
\item[\textit{{[}CC3{]}}] Conformance. All elements should conform to the property that is the subject of the certification.
\end{description}
The Certification Criteria can be translated into Specific Criteria (SC) per Product Area that indicate what formal, uniform, and conformant means for that Product Area.

As we are specifically interested in the requirements part of the SPCM we provide examples of specific criteria (SC) for the User Requirements product area with a [CC3] conformance property of `correctness and consistency':
\begin{description}[\IEEEsetlabelwidth{SC3.2}\IEEEusemathlabelsep]
\item[\textit{{[}SC1.1{]}}] Required Elements: functional requirements, nonfunctional requirements, glossary
\item[\textit{{[}SC2.1{]}}] Uniformity: all use cases follow the same template.
\item[\textit{{[}SC3.1{]}}] Manual checks on Correctness and Consistency:
\end{description}
\begin{itemize}
\item{No two requirements or use cases contradict each other;}
\item{No requirement is ambiguous;}
\item{Functional requirements specify what, not how;}
\item{Each requirement is testable;}
\item{Each requirement is uniquely identified;}
\item{Each requirement is atomic;}
\item{The glossary definitions are non-cyclic;}
\item{Use case diagrams correspond to use case text;}
\item{etc. (see~\cite{HeckLaQuSo} for a complete list).}
\end{itemize}

\subsection{A Verification Framework for Agile Requirements}
\begin{figure*}[!ht]
\centering
\includegraphics[width=5in]{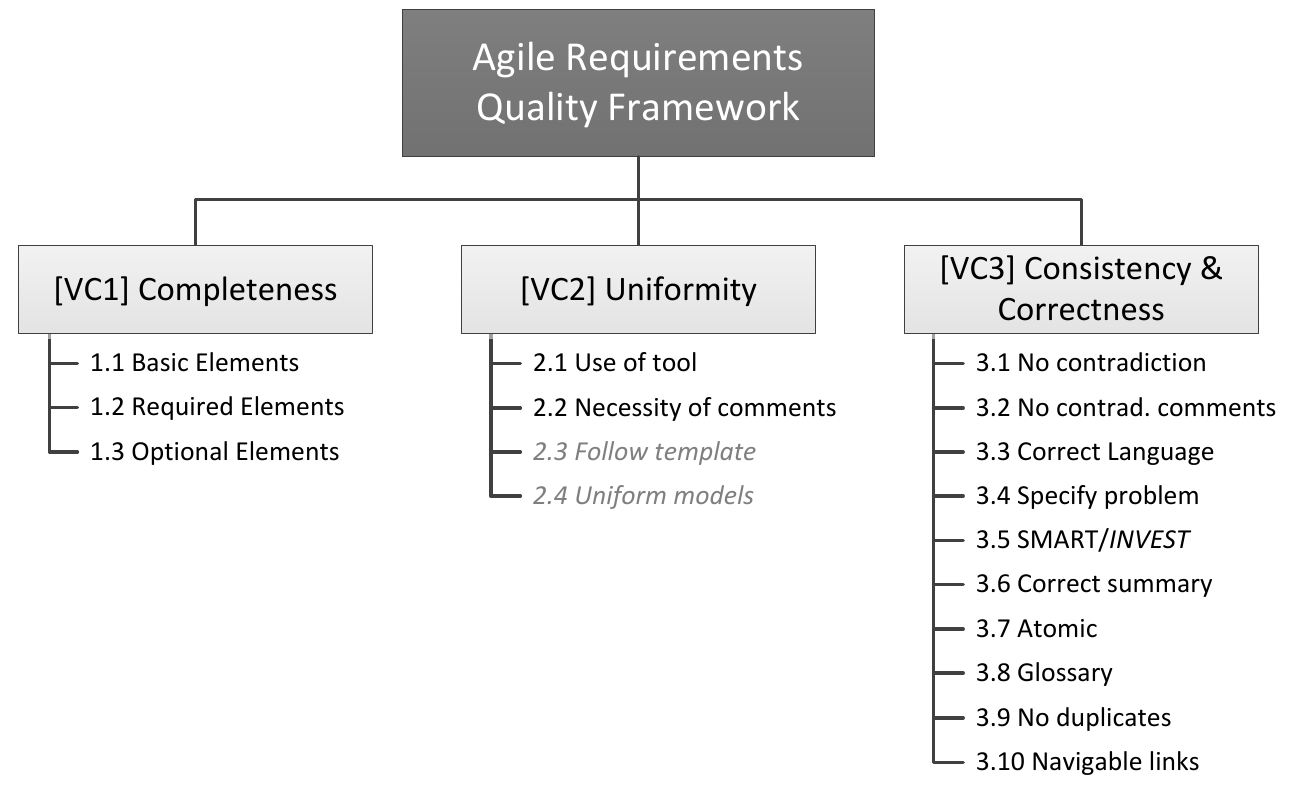}
\vspace{-2mm}
\caption*{\footnotesize{In \emph{italic} the items that only apply to user stories, the rest applies to both user stories and feature requests.}}
\caption{\textbf{Agile Requirements Verification Framework}}
\label{fig_1}
\vspace{-5mm}
\end{figure*}

Based on the SPCM we define the same three overall criteria for agile requirements. Because the purpose of our framework is not `certification' we rename them to Verification Criteria (VC):
\begin{LaTeXdescription}
\item[\textbf{{[}VC1{]}}] Completeness. All elements of the agile requirement should be present. We consider three levels: basic elements, required elements, optional elements. In that way we can differentiate between elements that are absolutely mandatory for a requirement and elements that are nice to have because they increase the requirement quality. 
\item[\textbf{{[}VC2{]}}] Uniformity. The style and format of the agile requirements should be standardized, because this leads to less time for understanding and managing the requirements. Each time a team member is confronted with a new requirement he/she needs some time to understand the requirement and decide what to do with it. This process takes less time when the requirements format is standardized. Then all team members know where to look for what information on the requirement or how to read certain models attached to the requirement. 
\item[\textbf{{[}VC3{]}}] Conformance. The agile requirements should be consistent and correct.
\end{LaTeXdescription}
The overall verification criteria are detailed into specific criteria [SCx.x] for each type of agile requirements. An overview of the framework can be seen in Figure~\ref{fig_1}. Figure~\ref{fig_1} includes an instantiation for user stories and feature requests, agile requirements types that are advocated by Ernst and Murphy~\cite{Ernst2012}.

\section{Feature Requests in Open Source Projects}
\label{FR}
\begin{figure}
\centering
\includegraphics[width=\columnwidth]{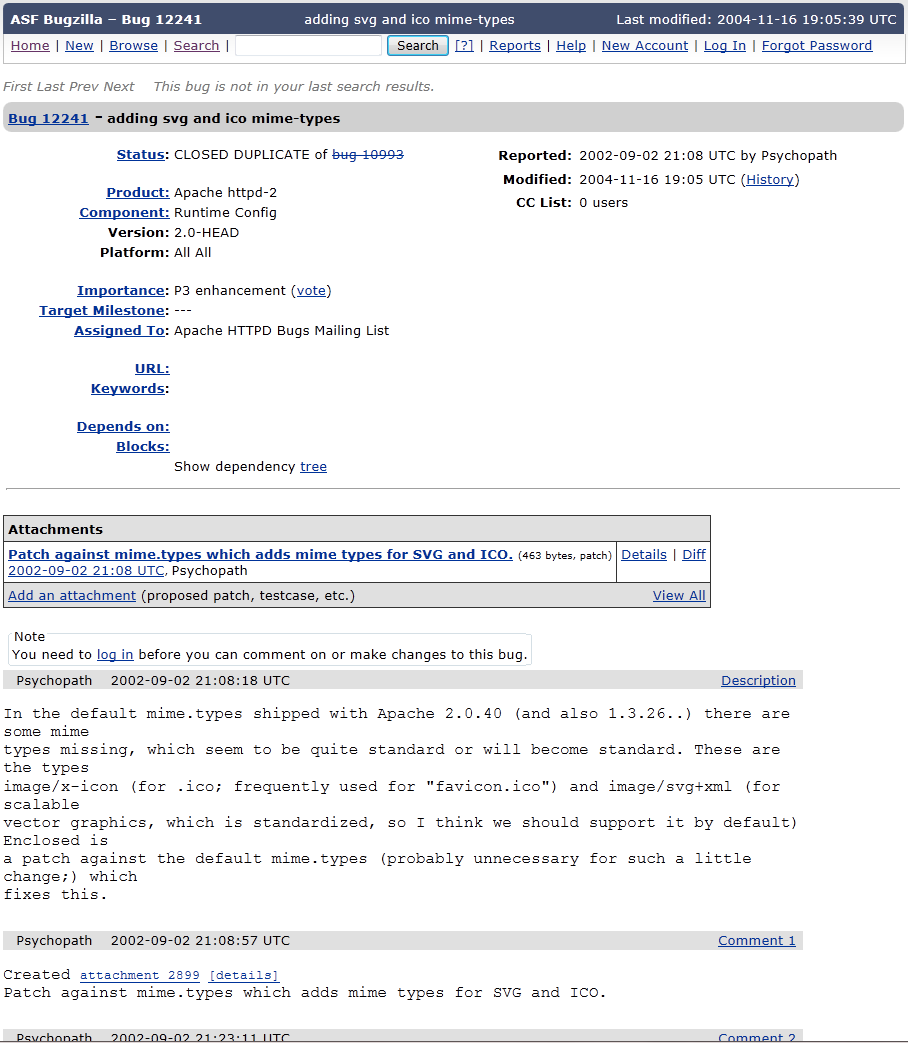}
\caption{\textbf{Feature Request in Bugzilla (HTTPD project)}}
\label{fig_bugzilla}
\vspace{-6mm}
\end{figure}

A feature request is a structured request (with a title, a description and a number of attributes) for new or enhanced functionality to an existing system. For an example, see Figure \ref{fig_bugzilla}.    

Previous work has shown that most open source projects use an issue tracker to manage feature requests~\cite{Heck2013}. In open source projects teams are usually spread around the planet without face-to-face contact~\cite{Warsta, Koch, MockusTOSEM2002}. This makes feature requests in open source projects more elaborate, as all communication needs to be documented on-line. User stories in agile projects document a minimum, because the team is usually co-located (even including the customer or a representative of the customer) and thus the main communication can be done off-line. The elaborate on-line documentation is why we take feature requests as our `base case'. 

At first sight, a feature request looks much like a traditional requirement (`a documented representation of a condition or capability needed by a user to solve a problem or achieve an objective'~\cite{IEEE}). In the below sections we analyze why and how feature requests are different from traditional requirements. The translation of this analysis to specific criteria for feature requests can be found in Table~\ref{table_model}. Not all criteria are relevant when a feature request is first specified. Some criteria, such as `link to source code' (SC3.1c) can only be fulfilled later in the life-cycle of the feature request. This means for some criteria we have to consider the complete feature request with all comments (added later by different team members). 

For the purpose of this paper, we focus on \emph{open source} feature requests because of their public availability. However, according to Alspaugh and Scacchi~\cite{Alspaugh} open source feature requests are very much like closed source feature requests, so most of our results should hold in both cases. 

\subsection{Completeness for Feature Requests}
\label{sec_prov}
Completeness (VC1) in our framework means that all elements of the specification are present. This should not be confused with the completeness of the content of the specification (`did we specify the complete user need?'). 

The SPCM (see section \ref{sec_SPCM}) considers a requirement specification complete if it includes use cases or functional requirements, behavioral properties (like business rules), objects (entity model or a glossary) and non-functional requirements. 

Alspaugh and Scacchi \cite{Alspaugh} find that the overwhelming majority of requirements-like artifacts in open source projects may be characterized as what they term \emph{provisionments}.  Provisionments state features in terms of the attributes provided by an existing software version, a competing product, or a prototype produced by a developer advocating the change it embodies. Most provisionments only suggest or hint at the behavior in question; the expectation seems to be that the audience for the provisionment is either already familiar with what is intended, or will play with the cited system and see the behavior in question firsthand.

This form of specification makes it hard to consider the completeness of elements in the same way as we do for traditional up-front requirements. We can however look at the attributes of a feature request. Which fields need to be filled for a feature request to be complete? The basic elements (SC1.1) are the ones that are inherently present in a feature request, such as title and description. The required elements (SC1.2) are the ones that are necessary for management of the feature request: keywords to organize them, a rationale to determine importance and a link to the source code once implemented for traceability. The optional elements (SC1.3) are the ones that add value to the developer when specified by the author, but can also be clarified in other ways (e.g. by prototyping or asking questions) later on in the process: scenarios, screen mock-ups or hints for a solution. 


\subsection{Uniformity for Feature Requests}
Uniformity (VC2) in our framework means that all requirements have the same format. 

For traditional up-front requirements, the SPCM~\cite{HeckLaQuSo} defines three levels of uniformity: all elements have the same format, all elements follow company standards, all elements follow industry standards. 

For feature requests in open source projects, company or industry standards usually do not apply. For example, feature requests are text-only, so no modeling language is used that can be compared to industry use. Most format choices for feature requests are determined by the issue tracker being used (SC2.1). Issue trackers have a number of pre-defined fields that must be filled in and that are always shown in the same way to users looking at the request (see for example Figure~\ref{fig_bugzilla}). 

The other thing to look at is the `uniformity of comments' (SC2.2). A feature request is entered with summary and description by the author. Then other persons (users or developers) can add comments to the feature request. This is done for discussion of the request or for tracking the implementation of the request. The comments in the different feature requests should be uniform, meaning they should be relevant for the given feature request. We define relevant as `necessary to understand the evolution of the feature request'. This is a subjective criterion but it definitely rules out comments like ``I am sorry for losing the votes.'' (Netbeans FR \#4619) or ``Wooo! Party :) thanks!'' (Netbeans FR \#186731).

\subsection{Correctness for Feature Requests}
\label{FR2}
Consistency and correctness (VC3) indicate those criteria that state something on the quality of an individual feature request (correctness), or on the quality of the link between two or more feature requests (consistency). 

In previous work we have distilled a list of quality criteria from literature and standards on requirements~\cite{HeckCSR}. This list is based on traditional up-front requirements. As we saw before not all elements from traditional requirements are relevant for feature requests (e.g., we have no complete object model). \new{The list for traditional up-front requirements as published in~\cite{HeckCSR} is:
\begin{compactenum}
\item{No two requirements contradict each other}
\item{No requirement is ambiguous}
\item{Functional requirements specify what, not how}
\item{Each requirement is testable}
\item{Each requirement is uniquely identified}
\item{Each use-case has a unique name}
\item{Each requirement is atomic}
\item{Ambiguity is explained in the glossary}
\end{compactenum}
}
For each item we will discuss if, and how, this applies to feature requests. \new{The resulting quality criteria for feature requests are mentioned between round brackets (SCx.y), see Figure \ref{fig_1} and Table \ref{table_model} for the description of those criteria.}\\
\emph{1)}
For a complete set of up-front requirements contradictions can more easily be established then for the ever-growing set of feature requests in an open source project. The feature requests are submitted by many different authors who often do not have a good picture of the feature requests that have been submitted before. However, the identification of related and possibly conflicting feature requests (SC3.1) is important for developers to determine the correct implementation. Another check that can be done is to see that the comments of a single feature request are not contradicting each other (SC3.2). \\
\emph{2)}
\new{As stated by Philippo et al.~\cite{Philippo} there are many factors that can decrease the effect of ambiguity and most of them are accounted for in agile environments.} For feature requests it is not such a big problem if the description is ambiguous because there is a habit of on-line discussion before implementation starts~\cite{Scacchi2009}. Another method that is frequently used in open source projects is prototyping~\cite{Alspaugh}. We can however require a basic level of clarity from the author of a feature request: write in full sentences without spelling/grammar mistakes (SC3.3). \\
\emph{3)}
As we saw above the author of a feature request may include hints for implementation of the feature request. As mentioned in~\cite{Noll2010} the majority of features is asserted by developers. This makes it more natural that some feature requests are stated in terms of the solution domain~\cite{Alspaugh}. They should however \emph{also} specify the problem that needs to be solved (SC3.4), for developers to be able to come up with alternative solutions.\\
\emph{4)}
As Alspaugh and Scacchi~\cite{Alspaugh} state an open source product that is evolving at a sufficiently rapid pace may be obtaining many of the benefits of problem-space requirements processes through solution-space development processes. This means that the fact that some feature requests may not be specified in a testable way can be compensated by extensive prototyping and involving the author of the feature request as a tester. However, we can require from the author to come up with verifiable feature requests and make the statement as precise as possible (SC3.5): e.g. ``I cannot read blue text on an orange background'' instead of ``I need more readable pages''.\\
\emph{5)}
A unique identifier is added automatically for each new feature request entered an issue tracker (IQ1, see Section \ref{IQ}). \\
\emph{6)}
Each feature request should have a unique name (`Summary' or `Title', SC1.1a). The summary should be in the same wording as the description and give a concise picture of the description (SC3.6).\\
\emph{7)}
For feature requests in an issue tracker it is very important that they are atomic, i.e. describe one need per feature request (SC3.7). If a feature request is not atomic the team runs into problems managing and implementing it (a feature request cannot be marked as `half done'). The risk also exists that only part of the feature request gets implemented because the comments only discuss that specific part and the other part gets forgotten.\\
\emph{8)}
In open source projects it is often assumed that users and developers involved are familiar with the terminology of the project (like `DnD' means `Drag and Drop') but the bigger and older the project gets, the more likely that new unfamiliar persons arrive. It is a good practice to maintain a glossary (wiki-pages maintained by the community can be a good solution) for such project-specific terms and abbreviations (SC3.8). The advantage of on-line tools is that one can easily link terms used in feature requests to such a glossary. 

From our own experience with open source projects \cite{Heck2013} we saw many duplicate entries in the issue trackers. This is a risk because discussions on both duplicate feature requests might deviate if the duplicate goes unnoticed (SC3.9). Worst case this leads to two different implementations of the same feature. Issue trackers offer functionality to mark feature request as `DUPLICATE' such that users and developers are always referred to the master discussion. 

A last item is about the linking of feature requests. Each link to another feature request should be clearly typed and navigable (SC3.10). If the author of a comment wants to refer to another feature request then he/she should make sure to insert a URL (some tools do this automatically when using a short-code like `\#\textless issue-id\textgreater') and to give a proper explanation why he/she is linking the two feature requests.


\section{User Stories}
\label{US}
As a feature request can also be described with one or more user stories~\cite{Leffingwell}, we investigate whether the same quality criteria should be upheld.  
A user story is the agile replacement for most of what has been traditionally expressed as a functional requirement statement (or use case). A user story is a brief statement of intent that describes something the system needs to do for the user. The user story takes a standard (user voice) form: `As a \textless role\textgreater, I can \textless activity\textgreater so that \textless business value\textgreater'~\cite{Leffingwell}.

For user stories most quality criteria are the same as for feature requests. Below in bold we detail the differences, based on~\cite{Leffingwell}. [SC1.x$^\prime$] indicates that the criterion has the same title as for feature requests, but with different elements that should be part of the user story. [SC2.3] and [SC2.4] are only valid for user stories, as can also be seen in Figure \ref{fig_1}. [SC3.5$^\prime$] is a special case as the SMART criterion~\cite{Doran}\footnote{SMART = Specific, Measurable, Assignable, Realistic and Time-related.} is deemed not to be appropriate for user stories (they specify an intent, not a detailed requirement). Leffingwell~\cite{Leffingwell} introduces INVEST~\cite{Wake} 
as the agile translation of SMART. The [SCx.x] that are not mentioned in the below list are valid for user stories without changes.
\begin{LaTeXdescription}
\item[\textbf{{[}SC1.1$^\prime${]}}] Basic Elements: Role, activity, business value (`Who needs what why?') instead of summary and description;
\item[\textbf{{[}SC1.2$^\prime${]}}] Required Elements: acceptance criteria or acceptance tests to verify the story instead of rationale (rationale is the same as business value in SC1.1$^\prime$);
\item[\textbf{{[}SC1.3$^\prime${]}}] Optional Elements: the team could agree to more detailed attachments to certain user stories (e.g. UML models) for higher quality;
\item[\textbf{{[}SC2.3{]}}] Stories Uniform: each user story follows the standard user voice form;
\item[\textbf{{[}SC2.4{]}}] Attachments Uniform: any modeling language used in the attachments is uniform and standardized;
\item[\textbf{{[}SC3.5$^\prime${]}}] INVEST: User stories should be Independent, Negotiable, Valuable, Estimable, Small, Testable~\cite{Wake}.
\end{LaTeXdescription}
Note that for user stories the inherent incompleteness and incorrectness of the specification is usually compensated by extensive informal communication between the stakeholders (represented by one single product owner) and the team members. A team working with user stories should therefore decide for themselves which of the quality criteria apply to their practice. If e.g. the product owner is in a remote location, then the quality criteria for documented user stories should be applied. If e.g. user stories are only documented as a `user voice statement' and comments are discussed off-line, then [SC2.2] and [SC3.2] do not apply. The quality criteria could be incorporated into the team's `Definition of Ready' \cite{Pichler} that determines when a user story is ready to be included in the planning for the next development iteration. 

\section{Inherent Qualities of Agile Requirements in Electronic Tools}
\label{IQ}
Most agile and open source projects use electronic tools to store the requirements (user stories or feature requests). In that case a number of quality criteria are automatically fulfilled. This is why we did not include them in our verification framework. We explain each of them briefly below. 
\begin{LaTeXdescription}
\item{[IQ1] Unique ID: as stated above an electronic tool will automatically assign a unique ID to each added requirement.}
\item{[IQ2] History: electronic tools automatically track all changes to a requirement. This can be viewed directly from the tool GUI or in the database. }
\item{[IQ3] Source: electronic tools automatically log the author of a requirement and the author of each comment.}
\item{[IQ4] Status: electronic tools have a separate `Status' field where the status of the requirement can easily be seen. Most tools support a work-flow in which the status field is updated (manually or automatically) based on the work-flow step the requirement is in.}
\item{[IQ5] Modifiable: electronically stored requirements are by definition modifiable \cite{Davis} because the tool provides a structure and style such that individual requirements can easily be changed. }
\item{[IQ6] Organized: electronic tools offer an easy way to add attributes to requirements. With built-in search options it is easy for the tool user to locate individual requirements or groups of requirements (e.g. for the same component or for the same keyword).}
\end{LaTeXdescription}
Teams that use the verification framework for their agile requirements should check if their tool also supports these six quality criteria by default. If not, it makes sense for them to include the not supported criteria as extra check in [VC1] or [VC3].

\section{Experimental Setup}
\label{sec_es}
For \new{an initial} validation of our framework we chose to interview eight practitioners from the Dutch agile community, sourced through our personal network. The interview consisted of two parts:
\begin{enumerate}
\item{General questions on agile requirements quality, including an exercise to evaluate feature requests from the Firefox (\url{www.mozilla.org/firefox}) project;}
\item{An exercise to use our quality model on feature requests from the Bugzilla (\url{www.bugzilla.org}) project, followed by several questions to rate the quality model.}
\end{enumerate}
The first part of the interview was done with minimal introduction from our side and above all without showing the participants our framework.

For the second part, we have turned our quality model into a checklist (in Microsoft Excel) for the participants to fill in. For each check the answer set was limited. When each check is filled in, the spreadsheet automatically calculates a score for each of the three specific criteria (SC) and an overall score for the quality of a single feature request (Low/Medium/High), see Section~\ref{sec_score} for the inner-workings.

The feature requests used for the exercise were manually selected by the first author using the following selection criteria: a substantial but not too big amount of comments (between 7 and 10) in the feature request, feature request has been implemented, contents of the feature request are not too technical (understandable for project outsiders). This last criterion is also why we selected the two projects: both Firefox and Bugzilla are well-known (types of) tools such that project outsiders should be able to understand or recognize the features. The feature requests were accessed on-line. 

The data sets (five feature requests from Firefox, ten from Bugzilla), Excel check list and interview questions (in Dutch) can be found on-line for reference~\cite{Figshare}.

\subsection{Scoring}
\label{sec_score}
The scoring algorithm has been specifically developed for the purpose of the experiment, to serve as a guidance for the practitioners on judging feature request quality. In Table \ref{table_model} it is indicated for each criterion what the answers to the check can be (column `Metric'). Based on the individual answers for each criterion, the spreadsheet calculates an overall score for a single feature request: `LOW', `MEDIUM' or `HIGH'.

For [SC1.1] till [SC1.3] the summary score is calculated by determining the percentage of subchecks that is answered with `Yes'. The overall score for [SC1] is calculated as `INCOMPLETE' when the score for [SC1.1] or [SC1.2] is below 100\%. If both [SC1.1] and [SC1.2] score 100\%, then the overall score for [SC1] is based on the score of [SC1.3]: `NORMAL' if 0\%, `BETTER' if 33$\frac{1}{3}$\%, `BEST' if equal to or higher than 66$\frac{2}{3}$\%. 

For [SC2] the overall score is identical to [SC2.2] because [SC2.1] is `Yes' for all feature requests in our experiment. Thus the outcome of [SC2] is a percentage. 

For [SC3] each subcriterion is translated into a percentage score. `Yes/No' is translated to 0\% or 100\%. The three-option answers are translated to 0\%, 50\% or 100\%. Then the overall score for [SC3] is calculated by taking the simple average of all percentage scores, because in our opinion not one subcriterion is more important for correctness than the other subcriteria.

For the final score we first look at the [SC1] score. If [SC1] scores `INCOMPLETE', the final score is always `LOW'. If [SC1] does not score `INCOMPLETE', the final score is a weighted average of [SC1.3], [SC2] and [SC3]. SC3 has a weight of 3 in this average as we feel that the `Correctness' is the most contributing factor to the overall quality of the feature request. [SC1.3] are `optional elements', [SC2] is mostly determined by the use of a tool (comments are just a small factor for uniformity) and [SC3] really looks at if everything that \emph{has} been written is written in a correct way. The overall quality score is considered `HIGH' when equal to or above 75\%, `LOW' when below 55\% and `MEDIUM' otherwise.    

This scoring algorithm is based on our professional opinion on what defines good quality. For other situations different rules or a (different) weighted average might be more appropriate. We decided to use this algorithm as a basis for the qualitative validation of our framework and plan to further investigate the scoring algorithm in a future quantitative study. 
\section{Interview Results}
\label{sec_res}
\subsection{Part One: Background}
The eight participants are experienced IT specialists with a good knowledge of the agile process. The participants originate from five different Dutch companies, that are active in both software development and quality consulting. All of the participants work in agile projects as coach, trainer or consultant. Most of them also have hands-on experience as analyst or tester in agile projects. All participants mention user stories as a format for agile requirements, but also use cases, features, and wireframes (i.e. screen blueprints) are mentioned. Some participants mentioned that they also consider informal communication as being part of `the agile requirement'. We made clear that for the purpose of our experiment we only consider the \emph{written} part.  

All participants agree that agile requirements should fulfill certain quality criteria. This helps the understanding within the team and is important for traceability or accountability towards the rest of the organization. When asked for a list of quality criteria the participants do not only mention verification criteria like the ones we have in our framework, but also include process-oriented criteria like ``have been approved by the product owner''. Most participants also emphasize the importance of validation (`is it what the user wants?') in agile projects.

When asked to score 2 feature requests from the Firefox project (175232 and 407117) as HIGH/MEDIUM/LOW quality (without prior knowledge of our framework, just based on professional opinion), the participants do not always agree on the exact score, but they consistently score 175232 lower than 407117.

\begin{figure}[t]
\centering
\includegraphics[width=0.9\columnwidth]{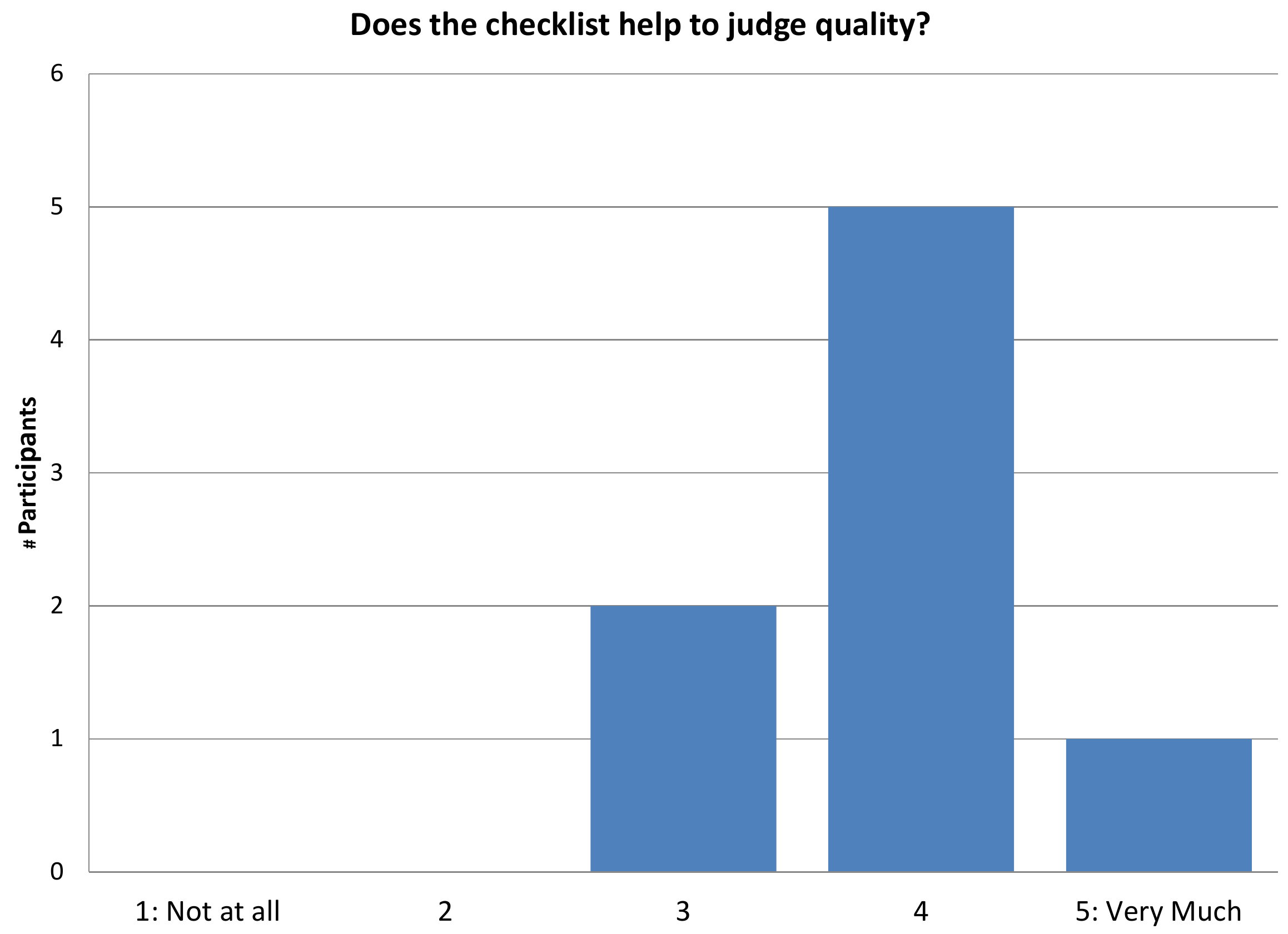}
\caption{\textbf{Participants rate the checklist as helpful}}
\label{fig_chart}
\vspace{-7mm}
\end{figure}

\subsection{Part Two: Our Agile Quality Framework}
We have asked each participant to fill in the checklist for at least two different feature requests from the Bugzilla project to get some hands-on experience with the checklist. The goal of this exercise was not to collect quantitative data, but to get qualitative feedback from the participants on the checklist. 

We learned that 4 participants mention ``\# of relevant comments'' (SC2.2) and 2 participants mention ``SMART'' (SC3.5) as checks that are unclear or difficult to fill in. For [SC2.2] they find it difficult to determine if a comment is `relevant' or not and for [SC3.5] they have difficulties determining the overall score on 5 criteria (Specific, Measurable, Acceptable, Realistic, Time-bound) in one check. We agree that these two checks are quite subjective, but we chose not to objectivize them in further detail. As one participant remarks: ``I am in favor of checklists but quantifying in too much detail triggers discussions on scores and weighing. The discussion should be on the content of the requirement.''. In practice the checks should not be used to get an exact score, but to get the professional opinion of the reviewer. As such it is much more important to find violations against those checks (e.g. ``Do I see not-SMART statements'' or ``Do I see irrelevant comments?'') and improve the requirement based on that.   

When asked to rate the score calculated by the Excel sheet for each feature request (LOW/MEDIUM/HIGH) the opinions vary. On a scale from 1 (no match at all with my personal opinion) to 5 (great match) all ratings have been given, although 6 out of 8 participants rate 3 or higher. This result indicates that most participants consider the final score of the model to be relevant. Yet, we also accept that our initial weighting scheme for the checklist requires fine-tuning in future experiments.
For example, in the checklist used in this experiment a feature request always scores LOW if one of the basic (SC1.1) or required (SC1.2) elements is missing and not all participants agree with this choice. They for example argue that a feature request with a missing `Rationale' (SC1.2b) or missing `Priority' (SC1.1c) can still be a correct feature request. In some situations for very simple and obvious feature requests it might be overkill to ask for all required elements, so then the feature request should still be able to score `HIGH'. We agree with this opinion. We added a scoring algorithm to help the participants in judging feature request quality, but the experiment also shows that the scoring algorithm should not be taken as an absolute judgement (one participant: ``A practical checklist like this always helps, but I am not sure how useful it is to calculate a final score from the individual checks.''). As stated before the checklist (as any checklist) is very useful as a reminder of what to check when looking for good feature request quality. It is the reviewer or author of the requirement that can still decide how serious a violation of one of the checks is in the given situation. The above example with the missing `Rationale' for obvious requirements could for instance be solved by marking this check as `N/A' in the Excel checklist. This would avoid the automatic `LOW' score. 

Some participants answered that they would like to add checks on more topics such as: uniformity of phrasing, hierarchy of requirements, non-functional impact (e.g. usability, performance), business value, approval from business/product owner, traceability, domain models. We see this as valuable suggestions for practitioners customizing the checklist for their own agile projects. We feel that none of these checks are feasible for feature requests in open source projects as we focused on in this experiment. As one participant mentions ``It is refreshing that this checklist is tailored for this specific situation. The ultimate result would be to know how to construct such a tailored checklist.''. The answer to this last question is simple in theory (the team decides on the specific checks and metrics for each of the three Verification Criteria) but at the same time difficult in practice (on what grounds would the team decide this?).  Our advice is to start with the checklist in Table \ref{table_model} and first decide which checks are not relevant for the given situation. The next step would be to see which checks should be added. These missing checks can be found by interviewing team members, by re-evaluating old requirements (why do we think this requirement is good/bad?), or by just applying the checklist in practice and improving it on-the-fly. As stated before, it might not be necessary to add metrics and overall scoring, so a simple Yes/No answer for each check with the goal to answer all checks positively could be enough for most teams.   

But why would teams do the effort of including such a checklist in their development process? All participants rated the checklist as helpful when judging the quality of a feature request (compared to part one of the experiment were they had to judge based on `gut-feeling’), see Figure~\ref{fig_chart}. They valued the help of the checklist to not forget criteria, to base their opinion on facts, to use it as an education for new team members, to standardize the verification process. One participant (that rated the checklist as `Very Much' helpful) nuances this by stating ``It is not always the case that high-quality requirements lead to high-quality products. The checklist is helpful but just a small part of all factors that influence final product quality.''. This is a valid point. Our study shows that also in agile environments requirements verification is considered important, but that there is no one-size-fits-all solution for agile teams. All participants confirmed that our framework is a good starting point to get to a tailored process for agile requirements verification.

\section{Discussion}
 \label{sec_dis}

In this section we first revisit the research questions, after which we discuss threats to validity.

\begin{table}
\centering
\begin{smaller}
\renewcommand{\tabcolsep}{0.1cm} 
\begin{tabular}{p{1.55in}|p{1.7in}}
\toprule
\textbf{Davis et al.} & \textbf{Agile Requirements Verification Framework} \\
\midrule
Unambiguous&[SC2.4], [SC3.3], [SC3.8]\\
Complete& [VC1]\\
Correct, i.e. contributes to the & [SC3.4], [SC1.2] - Rationale or \\
~~~ satisfaction of some need & ~~~ business value\\
Understandable& [SC1.3], [SC3.3]\\
Verifiable& [SC3.5]\\
Internally consistent& [SC3.1], [SC3.2], [SC3.6]\\
Externally consistent& N/A\\
Achievable& [SC3.5]\\
Concise& [SC2.2]\\
Design independent& [SC3.4] (Solution might be included)\\
Traceable i.e. facilitates referencing   & [SC3.7], [IQ1]\\
~~~ of individual requirements & \\
Modifiable i.e. table of contents  & [IQ5]\\
~~~ and index & \\
Electronically stored & [SC2.1]\\
Executable i.e. a dynamic behavioral & N/A\\
~~~ model can be made & \\
Annotated by relative importance& [SC1.1] - Relative importance\\
Annotated by relative stability& N/A\\
Annotated by version& [SC1.1] - Version\\
Not redundant& [SC3.9]\\
At right level of detail& [VC1]\\
Precise& [SC3.5]\\
Reusable& N/A\\
Traced i.e. clear origin & [SC1.2] - Link to code, [IQ2], [IQ3], [IQ4] \\
Organized& [SC1.2] - Keywords, [SC3.10], [IQ6] \\
Cross-referenced& [SC3.1], [SC3.9], [SC3.10]\\
\bottomrule
\end{tabular}
\end{smaller}
\caption{\textbf{Comparison between Davis et al. \cite{Davis} and our agile verification framework}}\label{table_SRS}
\vspace{-6mm}
\end{table}

\subsection{[RQ1a] Comparison to Traditional Requirements}
We started by asking: \emph{which quality criteria from literature on traditional up-front requirements do also apply to the verification of feature requests and user stories?} 
With regards to quality criteria for requirements in general, next to the aforementioned IEEE-830 standard~\cite{IEEE830}, a comprehensive work is the work by Davis et al.~\cite{Davis}, who performed a thorough analysis of qualities of a quality software requirements specification (SRS). Their analysis is based on an up-front requirements document. They have included metrics for each quality criterion. 

Consider Table~\ref{table_SRS} in which we analyze how each of the qualities identified by Davis et al.~\cite{Davis} for up-front requirements apply to agile requirements and where they have been included in our framework. As can be seen only four out of twenty-four criteria from Davis et al. are not incorporated into our agile verification framework:
\begin{itemize}
\item{Externally consistent. In both user stories and feature requests no external documents (Davis et al. define this as `already baselined project documentation') have to be considered. The so-called provisionments (see Section \ref{sec_prov}) are not specified with respect to other documents, but with respect to existing systems. Of course in specific situations external documents like a business case can be relevant. In that case the team should add one or more criteria to [VC1.3] to check the consistency between the agile requirement and the external document(s).}
\item{Executable. In agile and open source projects it is not common to spend that much effort on up-front specification. This is compensated by extensive prototyping or frequent releases.}
\item{Annotated by relative stability. Agile projects have embodied change as a known fact. All requirements can change. They solve this with short iterations and reprioritization of requirements for each iteration. That is why in agile projects we do not need a special attribute to specify change-proneness up-front.}
\item{Reusable. Since agile requirements are necessarily incomplete (`provisionments'), it makes no sense to reuse them.}
\end{itemize}

\subsection{[RQ1b] Additional Quality Criteria}
We subsequently asked ourselves whether \emph{we could see additional quality criteria for the verification of feature requests and user stories}?
All criteria from our resulting framework, see Table~\ref{table_model}, are in one way or another present in the work of Davis et al.~\cite{Davis}, see Table~\ref{table_SRS}. However, we have adjusted the specific description of each criterion to agile requirements, e.g. for the criterion `Design-independent' Davis et al. explain that a maximum number of designs should exist to satisfy user needs/external behavior. We have included `[SC3.4] - Specify Problem' but we specifically allow the user to \emph{also} specify design solutions, as this is common practice in open source projects where users are also developers.  

For user stories we have added one criterion: `follow the standard user voice form'. This is unique for user stories since traditional requirements and feature request do not follow standard templates (methods that advocate this are not widely used in practice). 

Overall we did not so much include `additional criteria', but we defined the `new' interpretations of existing criteria.

\subsection{[RQ2] Validation by Practitioners}
Finally, we wondered \emph{How do practitioners value our list of quality criteria with respect to usability, completeness and relevance for the verification of feature requests?}
The overall evaluation of the framework for open source feature requests was positive. The practitioners answers have made clear to us that specific situations need some fine-tuning of the specific checks and the scoring. Our framework caters for that kind of specific tailoring and we have given some hints om how to approach this. The feedback from practitioners did not make us change anything in our basic framework. 

\subsection{Threats to Validity}
A threat to validity in our experiment is the fact that we only had eight participants. However, the participants are from sufficiently different companies and backgrounds to get a first overall impression of community feedback. Furthermore we also validated our framework against existing literature in the area of requirements quality and the area of agile requirements.

Although we only personally knew two of them, the participants might have been inclined to rate our quality framework positively. We tried to mitigate this by informing all participants up-front that only honest answers were of use to us and we needed feedback on how to improve our quality framework.

\section{Related Work}\label{RelWork}
As mentioned in the introduction we did not find specific literature on the quality of agile requirements. The following publications are however somewhat related to our quality framework, so we discuss them briefly.

Dietze~\cite{Dietze} describes the agile requirements definition processes performed in open source software development projects. He describes the
typical requirements contribution and review processes and a typical life-cycle of the produced requirements artifact. He mentions the meta-data of a change request (corresponding to [SC1] in our framework), but does not discuss the aspect of change request quality.  

Scacchi~\cite{Scacchi2009} argues that  requirements validation is a by-product, rather than an explicit goal, of how open source software (OSS) requirements are constituted, described, discussed, cross-referenced, and hyperlinked to other informal descriptions of a system and its implementations. From his study it appears that OSS requirements artifacts might be assessed in terms of virtues like 1) encouragement of community building; 2) freedom of expression and multiplicity of expression; 3) readability and ease of navigation; 4) and implicit versus explicit structures for organizing, storing and sharing OSS requirements. Virtue 3) and 4) above are covered in our framework, whereas virtue 1) and 2) should be achieved by a correct setup of the open source project (allow everyone to report feature requests and provide good means and an open atmosphere for discussing them).

Bettenburg~\cite{Bettenburg} et al. have conducted a survey among developers and users of Apache, Eclipse, and Mozilla to find out what makes a good bug report. The analysis of the 466 responses revealed an information mismatch between what developers need and what users supply. Most developers consider steps to reproduce, stack traces, and test cases
as helpful, which are at the same time most difficult to provide for users. These three items can be compared to the scenario's and screens we have included in [SC1.3] for feature requests. 

G\'enova et al.~\cite{Genova} describe a framework to measure the quality of textual requirements. They have defined metrics and implemented those metrics in a tool to automatically verify the quality. A requirement is thus scored as being bad, medium or good. The tool is commercially available and users report benefiting from using it. G\'enova et al. use formal requirements documents as input data, making some of their quality criteria less relevant for agile requirements. See also our analysis of the earlier work of Davis et al.~\cite{Davis} in Section~\ref{sec_dis}. We do however value the idea of automating certain quality checks as was also requested by one of the participants in our experiment. This is something we plan to do in the future.  

\section{Conclusion} \label{Conclusion}
In this paper we have developed a quality framework for agile requirements that caters for tailoring to different situations. We have instantiated this framework for both feature requests (in open source projects) and user stories and given some hints how to do this for other situations. The framework is based on literature on the quality of traditional requirements as well as literature on agile requirements and open source projects. 

We have \new{performed an initial validation of} our framework with eight practitioners. The framework was positively evaluated by all of them. The practitioners also confirm our assumption that verification of agile requirements is important, the same as for traditional up-front requirements. 
In future work we plan to quantitatively evaluate the framework to see how `subjective' the checks are and deliver some more guidance on metrics and scoring. Another useful area for future work is automation of the checks as suggested by one of the participants in our experiment.

\new{It has been shown in the past that better quality of written requirements can contribute to a better overall quality of the end product. A next step is to verify this contribution with our framework in case studies in agile environments.}



\section*{Acknowledgments}
This work has been sponsored by the RAAK-PRO program under grants of the EQuA-project. We would like to thank all participants in the interview sessions: R. Wouterse (SYSQA), S. Jansen, H. Nieboer, P. Devick (Info Support), A. Grund (AGrund.informatiespecialist), S. van der Zee, A. Uittenbogaard (inspearit), R. van Solingen (TU Delft, Prowareness).



\begin{table*}[!t]
\centering
\begin{smaller}
\renewcommand{\tabcolsep}{0.1cm} 
\begin{tabular}{|c|p{3cm}|p{11.3cm}|p{2.2cm}|}
\multicolumn{1}{c}{\uppercase{\textbf{ID}}} & \multicolumn{1}{l}{\uppercase{\textbf{Criterion}}} & \multicolumn{1}{l}{\uppercase{\textbf{Description}}} & \multicolumn{1}{l}{\uppercase{\textbf{Metric}}}\\
\hline
\multicolumn{4}{|c|}{\rule{0pt}{1.5em} \cellcolor[gray]{0.75} \normalsize{\textbf{{[}VC1{]} Completeness}}}  \\[3pt]
\hline
\multicolumn{4}{|l|}{\cellcolor[gray]{0.9} \emph{SC1.1 Basic Elements}}  \\
\hline
\multicolumn{1}{|r|}{a} & Summary and Description & The `Description' contains the provisionment and the `Summary' (or `Title') gives a clear short version of the provisionment & yes/no\\
\hline
\multicolumn{1}{|r|}{b} & Product and Version & It should be indicated for which software product and version the provisionment holds & yes/no \\
\hline
\multicolumn{1}{|r|}{c} & Relative importance & The relative importance of the feature request should be clear. Examples of this are a `Priority' or `Severity' field or a voting mechanism (feature requests with more votes are more important) & yes/no\\
\hline
\multicolumn{4}{|l|}{\cellcolor[gray]{0.9} \emph{SC1.2 Required Elements}}  \\
\hline
\multicolumn{1}{|r|}{a} & Keywords/tags & For organization purposes (easily finding related requests) a feature request should be tagged with keywords & yes/no\\
\hline
\multicolumn{1}{|r|}{b} & Rationale & Each feature request should have a justification. The author should specify why this feature request is important for him/her. This helps the implementation team in deciding on the priority of the feature request & yes/no\\
\hline
\multicolumn{1}{|r|}{c} & Link to source code for fixed requirement & For solved feature requests it should be indicated in which version of the source code it has been solved. This can be done through a manual comment, or through an automated one generated from the source code management system. Ideally the feature request has a separate field to track the traceability to code & yes/no\\
\hline
\multicolumn{4}{|l|}{\cellcolor[gray]{0.9} \emph{SC1.3 Optional Elements}}  \\
\hline
\multicolumn{1}{|r|}{a} & Use case or Scenario & The author could specify the exact steps that he/she is missing in the current version of the software. What is the trigger for the missing functionality and what are the different scenario's in which the functionality would be useful? & yes/no\\
\hline
\multicolumn{1}{|r|}{b} & Screens & The author could clarify the screens in the existing system that he/she wants to be changed by adding screen shots of the current situation and/or screen mock-ups of the desired situation. & yes/no\\
\hline
\multicolumn{1}{|r|}{c} & Possible solution &  The author could add a complete solution as `Attachment' (patch), but could also specify hints for a possible solution in comments & yes/no\\
\hline
\multicolumn{4}{|c|}{\rule{0pt}{1.5em} \cellcolor[gray]{0.75}  \normalsize{\textbf{{[}VC2{]} Uniformity}}} \\[3pt]
\hline
SC2.1 & Issue tracker or other tool should be used&  The use of an issue tracker ensures that feature requests are stored in an uniform way, at least with respect to the attributes that are present for the feature request. Of course it remains up to the author to correctly fill those fields &yes/no\\
\hline
SC2.2 & All comments are necessary &  All comments are necessary to understand the evolution of the feature request. The addition of irrelevant comments (e.g. thank you notes or 
instructions how to behave in the project) lowers the quality of the feature request and makes it more difficult for people to understand the totality of it 
& The percentage of relevant comments\\
\hline
\multicolumn{4}{|c|}{\rule{0pt}{1.5em} \cellcolor[gray]{0.75} \normalsize{\textbf{{[}VC3{]} Consistency and Correctness}}} \\[3pt]
\hline
SC3.1 & No contradicting feature requests & It is difficult to completely avoid conflicting requests, but they should never go unnoticed. A link can be made through comments and one of the feature requests should be `Closed' to avoid implementation of the wrong request & yes/no \\
\hline
SC3.2 & No contradicting comments & Each contradicting comment should be clarified in later comments. It should be clear what the correct interpretation of contradicting comments is & A lot, a few, none\\
\hline
SC3.3 & Correct language & Feature requests should be written in full sentences without hindering spelling or typing errors & Very much, a little bit, not at all \\
\hline
SC3.4 & Specify problem & Feature requests may include (hints for) a solution, but should always describe the problem that needs to be solved. This helps developers to think about alternative solutions &  yes/no\\
\hline
SC3.5 & SMART & A feature request can be more quickly and easily resolved if it is Specific, Measurable, Acceptable, Realistic and Time-bounded (SMART \cite{Doran}) & Very much, a little bit, not at all \\
\hline
SC3.6 & Correct summary & The summary should be a brief statement of the needed feature. It should be clear from the summary what the feature request is about. The description should give added value to the summary. 
& yes/no\\
\hline
SC3.7 & Atomic & Each feature request should contain only one requirement &  yes/no\\
\hline
SC3.8 & Glossary & Each unclear term or abbreviation should be explained in the feature request or in a separate `glossary' & Very much, a little bit, not at all \\
\hline
SC3.9 & No duplicate requests & Having too many duplicate requests clutters the database. Due to the nature of open source projects there will always be some duplicates (also users not so familiar with the project get the rights to enter requests). At least the duplicates should be identified, properly linked, and the master should be indicated. 
&  yes/no\\
\hline
SC3.10 & Navigable links & Links to other feature requests should be navigable (by clicking on the link) and it should be explained what the type of the link is & yes/no\\
\hline
\end{tabular}
\end{smaller}
\vspace{-0.1cm}
\caption{\textbf{The Agile Quality Framework for Feature Requests}}
\label{table_model}
\vspace{-0.5cm}
\end{table*}

\bibliographystyle{IEEEtran}
\bibliography{CSR2014Heck}

\end{document}